\documentstyle[a4wide, epsfig]{article}
\title{Quantum Principles and Mathematical Computability}
\author{Tien D Kieu\footnote{email: kieu@swin.edu.au},\\
Centre for Atom Optics and Ultrafast Spectroscopy,\\Swinburne
University of Technology, Hawthorn 3122, Australia}
\begin{document}
\maketitle
\begin{abstract}
Taking the view that computation is after all physical, we 
argue that physics, particularly quantum physics, could help extend
the notion of computability.  Here, we list the important and unique
features of quantum mechanics and then outline a 
quantum mechanical ``algorithm" for one of the insoluble problems 
of mathematics, the Hilbert's tenth and equivalently the
Turing halting problem.  
The key element of this algorithm is the {\em computability} and
{\em measurability} of both the
values of physical observables and of the quantum-mechanical 
probability distributions for these values.
\end{abstract}
\begin{flushright}
{\em The fact is that quantum computers can prove theorems by methods that 
neither a human brain nor any other Turing-computational arbiter will ever
be able to reproduce.  What if a quantum algorithm delivered a theorem that it
was infeasible to prove classically.  No such algorithm is yet known, but nor is anything
known to rule out such a possibility, and this raises a question of principle:
should we still accept such a theorem as undoubtedly proved?  We believe that the rational
answer ot this question is yes, for our confidence in quantum proofs rests upon the same 
foundation as our confidence in classical proofs:  our acceptance of the physical laws 
underlying the computing operations.}\\
D. Deustch, A. Ekert and R. Lupacchini~\cite{DEL}
\end{flushright}
\section{Introduction}
Supported by the convergence of many seemingly different models of
computation put forward independently by Turing, Post, Markov and 
others~\cite{computation}, a thesis on the notion of 
computatbility has been formed and gained much credibility.  The Church-
Turing thesis which can be phrased as
\begin{quotation}
\em Every function which would naturally be regarded as computable can be
computed by a universal Turing machine.
\end{quotation}
This is neither a theorem nor a conjecture, for it is not and 
cannot even be hoped to be proven.  The thesis simply asserts some correspondence
between certain informal concept, that of computability in this case,
with certain logically well-defined (i.e. mathematical) object, namely, 
the universal Turing machine.  

The thesis thus imposes an upper limit of what any computing machine
can be designed to do.  Can this computability notion be enlarged?   In principles, there is no
reason why not.  Proposals to overcome the Turing-machine limit range from
the models of mathematical principles such as continous valued neural
networks~\cite{sigelmann}, DNA computing~\cite{DNA} to those of physical
nature based on general arguments~\cite{stannett}, relativity
principles, and quantum mechanical principles~\cite{kieu1, kieu2, calude}.  

We summarise a quantum computing model in this paper.  
But first we present the quantum principles in the next section.

%


\section{Quantum principles}
So, what are the extra-logical features of Quantum Mechanics that would
enable an enlargement of computability?  Following Feynman, we will employ the gedanken ``simple" two-slit 
experiment, Figure 1, to point out all that can
and cannot be known about, but will be manifest in the weird reality of quantum 
physics.  This thought experiment is about a
plane wave of electrons -- all of the electrons in which have a single, well-defined 
value for the momentum -- passing through two slits one
by one before arriving at a detection screen where each electron can be
recorded at a definite position on the screen and at a definite moment in
(laboratory) time.
\begin{figure}
\begin{center}
\psfig{file=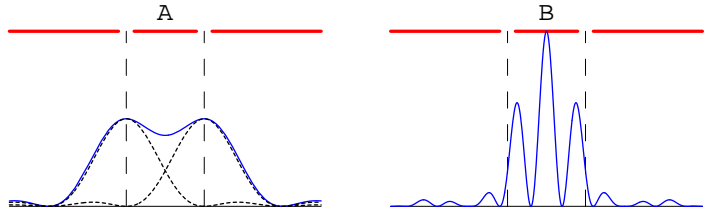}
\caption{Plane-wave electrons passing through the two slits from the top to arrive 
at the screen at the bottom: (A) The intensity pattern on the screen shows no
interference (continuous curve) when there is some mechanism to detect 
which slit the electrons have passed through (dashed curves are the intensities 
obtained when the other slit is closed); 
(B) Interference is clearly exhibited in the intensity pattern on the screen
when no record is kept of which slit the electrons have passed through.}
\end{center}
\label{figone}
\end{figure}

\subsection{Intrinsic randomness}
One important property of Quantum Mechanics is the randomness in 
the outcome of a quantum measurement.  Even if we prepare the initial
quantum states to be {\em exactly} the same in principle, 
(say, the plane-wave state for the electrons) we can still have 
different and random outcomes in subsequent measurements (like finding 
out which slit of the two that an electron so prepared would go
through, or where the electron would land on the final screen).  
Such randomness is a fact of life in the quantum reality of
our universe.

To reflect that intrinsic and inevitable randomness, the best that Quantum
Mechanics, as a physical theory of nature, can do is to list, given the 
initial conditions, the possible values for measured quantities and the
probability distributions for those values.  Both the values and the
probability distributions are {\em computable} in the sense that they
can be evaluated algorithmically to any desirable
accuracy~\cite{computable}{\footnote {More precisely mathematically, the
wavefunctions generated by unitary (hence, bounded) time-evolution operators are computable,
and also are
the eigenvalues of hermitean operators corresponding to measured observable values.  
See~\cite{pour-el} for more details.}}.  This
definition of computability of a number when it can be evaluated to any
degree of accuracy is sufficient to interpret the number and to establish 
its relationship with other numbers.

On the other hand, not only the values registered in the measurement of
some measurable but also the associated probability distributions
are {\em measurable} in the sense that they can be obtained to any
desirable accuracy by the act of physical measurements~\cite{computable}.  
Normally, the
values for measurable are quantised so they can be obtained exactly; the
probability distributions are real numbers but can be obtained to any
given accuracy by repeating the measurements again and again 
(each time from the same initial quantum state) until the 
desired statistics can be reached.  That is how the computable numbers from Quantum Mechanics can be judged
against the measurable numbers obtained from physical experiments.  Thus far, 
there is no evidence of any discrepancy between theory and experiments.


\subsection{Implied infinity}
Randomness is, by mathematical definition, incompressible and irreducible.  
In Algorithmic Information Theory, Chaitin~\cite{chaitin} defines 
randomness by program-size complexity: a binary string is considered 
random when the size of the shortest program that generates that string 
is not ``smaller" than the size of the string itself.  We refer the readers 
to the original literature for more technically precise definitions for 
the cases of finite and infinite strings.

Another way to see that randomness does entail infinity is given by an
interesting argument by Stannett~\cite{stannett} based on K\"onig's Lemma 
which states that any finitely-branching tree which contains infinitely 
many terminal nodes must also contain an infinite path.
It is pointed out that a classical algorithm which could generate
a truly random binary sequence must contain infinitely many terminal nodes 
(c.f. the program-size definition for randomness).  If this algorithm is 
recursive then it is possible that it may run forever without halting (that is,
along the infinite path enabled by the K\"onig's Lemma) 
in the generation of some single digit of the sequence.  As such, the 
algorithm does not really exist, for it cannot really generate a sequence if it
is stuck indefinitely at the generation of some intermediate bit
somewhere in the sequence. 

In sharp contrast, we can exploit Nature to 
generate an infinite binary sequence which is 
random just by, say, repeatedly detecting which of the two slits 
(hence the binary valuedness) the plane-wave electron goes through one by 
one\footnote{This possibility might seem to be unremarkable when compared 
to the tossing of a fabled, unbiased coin until it is pointed out that 
such a coin cannot exist classically.  The same conditions for tossing, 
i.e. the same initial conditions, will result in the same outcomes as 
required by the equations of classical physics.}.

Thus, paradoxically, the quantum reality of Nature somehow allows us 
to {\em compress} the {\em infinitely incompressible} randomness into the 
{\em apparently finite} act of preparing the same quantum state over and over again
for subsequent measurements!\footnote{What is finite here is the time 
taken to generate each and every digit of the sequence.}

Infiniteness implied by randomness is not, however, the only implied infinity 
that is embraced by quantum reality.  Quantum Mechanics suggests that an electron
would explore an infinite number of paths in going from one point to
another (say, from one slit to a point on the screen, in which case an infinity
is somehow ``contained" in the {\em finite} distance between the slit and the 
screen!).  This infinite multiplicity of the paths taken forms the basis
for Feynmann path integral formulation of Quantum Mechanics.  An infinity
within the finite would normally entail inconsistency -- or so one would deduce from
mathematical logic.  Amazingly, quantum reality manages to maintain the
required consistency by changing the outcome of the measurement as soon as we
try to detect/confirm the infinity by identifying the paths taken in 
between the finite separation.  This can be illustrated by and is in fact 
born out in the entirely different pattern (of no interference, see (A) in 
Fig. 1.) which will be recorded on the screen if we try to detect which of 
the two slits the electrons have gone through on their way there.  
The quantum mechanically implied infinity is both needed for and consistent
with the finitely measured!


\subsection{Quantum logic}
The above peculiar properties can be captured and manisfest in the
peculiarity of Quantum Logic.  In contrast to the classical logic of propositional 
calculus, quantum propositions $A$, $B$ and $C$ (each has a value of being either TRUE or FALSE)
do not in general satisfy the distributivity or modularity
property~\cite{qulogic}; that is,
\begin{eqnarray}
A\wedge(B\vee C) \not= (A\wedge B)\vee(A\wedge C).
\label{qulogic}
\end{eqnarray}

An heuristic example of this inequality can also be found in the
gedanken two-slit experiment if we label the propositions as
\begin{itemize}
\item $A$: the detection of an electron at a position $x$ on the final
screen;
\item $B$: the detection of an electron passing through one particular slit 
(say, the left one in Fig. 1) on its way to the final screen;
\item $C$: the detection of an electron passing through the other slit
on its way to the final screen.
\end{itemize}

With these choices and with the assumption that the electron intensity 
is low enough such that on average the electrons arrive at the screen one 
by one (i.e. no coincidence in the detection), the lhs of~(\ref{qulogic}) 
represents the detection of an electron at a position $x$ on the screen 
without ever knowing which slit it has gone through to get there.
On the other hand, the rhs of~(\ref{qulogic}) represents the detection 
of an electron at a position $x$ on the screen when it is known which 
slit of the two it {\it definitely} has passed through on its way there.
The former case experimentally gives rise to an inteference pattern on 
the screen, built up by the electrons one by one as in (B) of Fig .1.  
The latter gives rise to a non-interference pattern as in (A) of Fig .1, 
and thus is distinguishable from the former.
In particular, we can find a position $x$ on the screen where no electron 
is ever detected if we have interference (that is, at the node of 
interference pattern).  For this position $x$ the truth value of the 
combined proposition on the lhs of~(\ref{qulogic}) is thus FALSE; while 
that of the rhs is TRUE.  We then have the inequality in~(\ref{qulogic}).

\section{A quantum algorithm}
We will consider the equivalent
Hilbert's tenth problem~\cite{hilbert10th, davis}.  This problem, 
appropriately rephrased, asks for a general algorithm to determine if any 
given Diophantine equation has a (non-negative) integer solution or not.  
A Diophantine equation involves polynomial equation of many unknowns and 
integer coefficients.  If we can find a general algorithm asked for 
by the Hilbert's tenth problem then we will have a general algorithm 
for the well-known Turing halting problem; that is, we will be able to tell if any 
given Turing program will halt or not upon starting with some input.

Classically, there is no such algorithm by the Cantor's diagonal
arguments.  For this particular
Hilbert's tenth problem, we can see that its noncomputability 
originates from the lack of a general method to verify a
negative statement concerning solution of a Diophantine equation.  By 
direct subsitution into the Diophantine polynomial, it is
straightforward to verify whether a set of integers is indeed a zero of
the polynomial as it is claimed or not.  But substitution can not be used
to verify {\it in general} a negative statement that a 
Diophantine polynomial has no zero as it would require the infinite task of substituting {\em all}
integers!  For a particular equation, such as
the Diophantine equation of the Fermat's last theorem, one may be able
to find a specific way to confirm that the equation has no solution.
But that specific way is peculiar and only applicable to the equation in
consideration, or some related equations, and not to {\it any}
Diophantine equations in general.  

Nevertheless, a quantum algorithm has been proposed
recently~\cite{kieu1, kieu2} for the Hilbert's tenth problem.  
We will summarise the main points of the algorithm below and only wish 
to mention here that we consider quantum algorithms are as good as any 
algorithms in the sense that they can be implementable in the physical world, 
occupying finite time duration and finite spatial extent and consuming finite 
physical resources.

\subsection{Outlines}
Our strategy is that we do not look for the zeroes of the Diophantine 
polynomial in question, which may not exist, but instead search within the 
domain of non-negative integers for the 
absolute minimum of the square of the polynomial, which always exists 
and is finite.  While it is equally hard to
find either the zeroes or the absolute minimum in classical computation, 
we have converted the problem to the realisation of the ground state of
a quantum Hamiltonian and there is no known quantum principle against
such an act.  In fact, there is no known physical principles 
against it.  Let us consider the three laws of thermodynamics concerning 
energy conservation, entropy of closed systems and the unattainability
of absolute zero temperature.  The energy involved in our algorithm
is finite, being the ground state energy of some Hamiltonian.  The
entropy increase which ultimately connects to decoherence effects is a
technical problem for all quantum computation in general.

It may appear that even the quantum process can only explore a
finite domain in a finite time and is thus no better than a classical
machine in terms of computability.  But there is a crucial
difference.

In a classical search, even if the global minimum is encountered, it cannot 
generally be 
proved that it is the global minimum (unless it is a zero of the Diophantine 
equation).  Armed only with classical logic, we would still have to 
compare it with all other numbers from the 
infinite domain yet to come, but we obviously can never complete this 
comparison in finite time -- thus, the noncomputability.

In the quantum case, the global minimum is encoded in the energy of the ground
state of a suitable Hamiltonian.  Then, by energetic tagging, the global 
minimum can be found in finite time and confirmed, if it is the ground state that is obtained at the 
end of the computation.  
It is the physical principles that can be utilised to identify and/or verify the ground state. 
These principles are over and above 
the mathematics which govern the logic of a classical machine and help
differentiate the quantum from the classical.  Quantum mechanics could 
``explore" an infinite domain, but only in the sense that it can select, 
among an infinite number of states, one single state (or a subspace in case 
of degeneracy) to be identified as the ground state of some given Hamiltonian
(which is bounded from below).  

Our proposal is in
contrast to the claim in~\cite{QTM} that quantum Turing machines
compute exactly the same class of functions as do Turing machines, albeit 
perhaps more efficiently.  The quantum Turing machine approach considered there is a
direct generalisation of that of the classical Turing machines but with qubits
and some universal set of one-qubit and two-qubit unitary gates to build up,
step by step, dimensionally larger, but still dimensionally finite unitary operations.  
This universal set is chosen on its
ability to evaluate any desirable classical logic function.
Our approach, on the other hand, is from the start 
based on infinite-dimension Hamiltonians
and also based on the special properties and unique status of their ground states.  
The unitary 
operations are then followed as the Schr\"odinger time evolutions.

\subsection{Verification of the ground state}
The quantum algorithm is based on the key ingredients of:
\begin{itemize}
\item  The exactitude, to the level required, of the theory of Quantum
Mechanics in describing and predicting physical processes.
\item  Our ability to physically implement certain Hamiltonians having infinite
numbers of energy levels;
\item  Our ability to physically obtain and verify some state as the
desirable ground state;
\end{itemize}
If any of these is not achievable or approximable with controllable
accuracy, the quantum algorithm simply fails and further modifications 
may or may not work.

Without any known physical principles outlawing these key assumptions,
we sketch here an approach to obtain and verify the desirable ground state of
the Hamiltonian corresponding to the Diophantine polynomial in consideration.

It is in general easier to implement a hamiltonian $H_P$
than to obtain its ground state $|g\rangle$.  We thus should start the
computation in yet a different and readily obtainable initial state
$|g_I\rangle$, which is the ground state of some other 
hamiltonian, $H_I$, then deform this hamiltonian $H_I$ adiabatically in time
into the hamiltonian whose ground state is the desired one, through a 
time-dependent process represented by an interpolating Hamiltonian 
${\cal H}(s) = (1-s) H_I + sH_P$, for $s$ changes from 0 to 1.  The theorem of
quantum adiabtic processes ensures that we can get arbitrarily close to the
ground state $|g\rangle$ of $H_P$.  Figures 2 and 3 below give an heuristic 
illustration of the quantum adiabatic theorem, which can be exploited for 
optimisation problems.

\begin{figure}
\begin{center}
\psfig{file=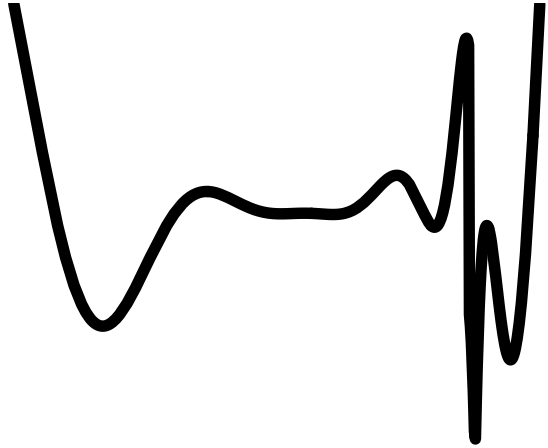}
\caption{An example of a landscape with a small basin attraction for the global minimum. 
Finding the quantum mechanical ground state of such landscape is quite difficult.}
\end{center}
\label{fig2}
\end{figure}

\begin{figure}
\begin{center}
\psfig{file=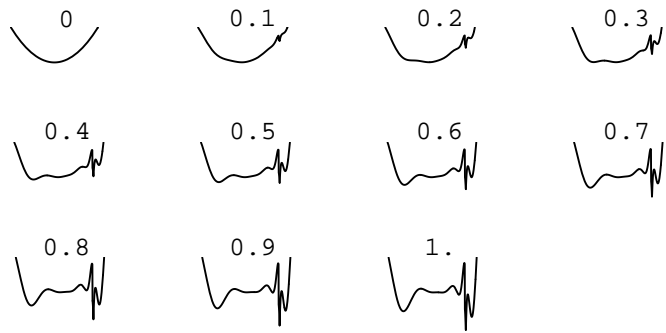}
\caption{Exploiting Quantum Adiabatic Theorem, we start with a simple landscape at the reduced time $s=0$ 
then adiabatically change it to the landscape in Figure 2 at $s=1$.  As time evolves, the readily available 
initial ground state quantum mechanically tunnels into the sought-after ground state of the landscape of Figure 2.}
\end{center}
\label{fig3}
\end{figure}

In order to solve the Hilbert's tenth problem we need on the one hand such time-dependent physical 
(adiabatic) processes to arrive at a candidate state.  On the other hand, the theory of Quantum
Mechanics can be used to verify whether this candidate is the ground state through the usual
statistical predictions from the Schr\"odinger equation with a truncated number of energy 
states of the time-dependent Hamiltonian ${\cal H}(s)$.  This way, we can overcome the 
problem of which states are to be included in the truncated basis for a numerical study 
of Quantum Mechanics.  This also reconciles with the Cantor diagonal arguments which state that 
the problem could not be solved entirely in the framework of classical computation. 

The key factor in the ground state verification is {\em the probability distributions}, 
which are {\em both} computable from a numerical study of Quantum Mechanics (that is,
with the control in the calculation to reach any desirable accuracy) {\em and} measurable 
in practice (i.e., by repeating the physical processes to obtain the statistics to any 
desirable accuracy).  By matching the calculated with the measured, both of which depend on the
evolution time which we can vary, we then
can unambiguously identify the ground state of the final Hamiltonian.
The information about the existence of solution, or lack of it, for
the given Diophantine polynomial can be inferred through some further quantum
measurements on this ground state.

It is worth noting that we have here a peculiar situation in which the computational 
complexity, that is, the evolution time, might not be known exactly {\em before}
carrying out the quantum computation -- although it can be estimated approximately.

\section{When is a proof a proof?}
Proof, be it mathematical or general, is the means to an end: a proof 
is there to explain, convince or persuade the others (and even oneself)
the ``truthful" value of certain statement(s).  

A classical proof, based on classical logic, starts with a finite number
of axioms from which a finite number of subsequent/intermediate
statements can be derived with the help of a finite number of inference
rules.  And it ends with a final statement, the ``truth".  All these
finiteness requirements are there to ensure the reproducibility of the
proof in a finite time and manner.  A classical algorithm is a
particular case of such a type of proof, with input being (part of) 
the axioms and output the final ``truth".

Deutsch and some others~\cite{DEL, virtual} see such a classical 
proof/algorithm as an {\em object}, static with intermediate records.  
On the other hand and in contrast, quantum algorithms are seen as dynamical processes 
wherein the intermediate ``steps" cannot be recorded without destroying 
the interference and thus the algorithms themselves.  The intermediate 
quantum ``steps", as a matter of facts, cannot be even made out as clearly 
defined, discrete steps, expressible in terms of classical propositions.

The requirement of finiteness of the number of intermediate steps in a
classical proof is no longer relevant (and unobtainable in a quantum
proof anyway) if consistency is maintainable and reproducibility is
achievable in quantum ``proofs".  

For such quantum proofs to be credible,
they must be consistent: a statement and its negation cannot be proved
at the same time under the same conditions.  In basing the notion of proof on 
the physical reality, the condition of consistency should be automatically 
and implicitly guaranteed, for after all there is only one reality -- 
or at most one which we can perceive.  Signs of inconsistency would not 
be pointing to something intrinsic of the
reality, but would be only because of our perception or understanding 
of reality.  In other words, inconsistency, as we see it, cannot refute 
the reality; it only hints that it is time we need a new theory of
Nature, which in turn may or may not affect the ``proof."

For such quantum proofs to be of some usefulness, they must be repoducible: 
reproducible in a finite duration of time, reproducible in at different 
locations and reproducible at different times.  The latter two requirements 
are ensured by the principles of invariance under spatial and temporal 
translations, which are some of the most cherished physical principles.  
These principles can be tested, and have been tested extensively to the 
hisghest accuracy without any failures, via their consequences in the 
conservation of energy and linear momentum.

But can they, the quantum proofs, be acceptable as proofs?  We would prefer 
an affirmative answer even though the answer to this question might only be 
a matter of taste.

It should be easier to accept a quantum process a proof if the end result 
of the quantum process can be verified by classical means -- for instance, 
in a factoring problem, the obtained primes can be easily multiplied together
to give back the original number as a check.  Similarly, a quantum process
should also be acceptable as a valid proof even if such {\em direct} 
verification cannot be carried out as a matter of principle 
but the result somehow can be verified indirectly through some other (physical or mathematical) 
handles.  This is the case of our algorithm for Hilbert's tenth problem 
above where we only need to verify that some state is indeed the ground 
state, a physical attribute only indirectly linked to the mathematical 
solutions sought.  

The authors whose quotation is quoted at the beginning of this article
have also argued that even when, as a matter of principles,
there is no direct nor indirect verifications possible, quantum process should
still be considered as valid means of proof, simply because of
``our acceptance of the physical laws underlying the computing operations."

\section{Concluding remarks}
In this paper we review the important characters of Quantum Principles and 
put forward the arguments that these physical principles
may help compute some of the classically noncomputable.  We also outline 
a recently proposed quantum algorithm for the Hilbert's
tenth problem, and emphasise the key role of probability distributions 
in the crucial verification, of a physical nature, for the solution for
this classically noncomputable problem. 

It remains to be seen if and when the quantum algorithm can be
physically realised.  If not prohibited by any physical principles, 
and we know none so far, then we trust that it can be implementable 
and will be realisable.  Whatever the case it may turn out to be, our
investigation has already opened up new and interesting directions for
Mathematics itself.  Our quantum algorithm has inspired a reformulation
of the Hilbert's tenth problem, a problem in the domain of the discrete
integers, in terms of a set of infinitely coupled differential
equations over continuous variables~\cite{kieu3}.  This may lead to 
new insights and/or solution of the problem (recalled that, despite the
mathematical noncomputability of Hilbert's tenth problem, there does
exist a general procedure to decide whether any given polynomial with
many unknowns and {\em real} (continuous) coefficients has {\em real}
solutions or not~\cite{tarski}.)

Our decidability study here in the framework 
of Quantum Mechanics only deals with the 
property of being Diophantine, which does not cover the property of being arithmetic in 
general (which could involve unbounded universal quantifiers).  This entails 
that our consideration has no direct consequences on G\"odel's Incompleteness theorem.

\section*{Acknowledgement}
I am indebted to Alan Head and Peter Hannaford for encouragement and
discussions; and would like to acknowledge the discussions with
Cristian Calude, Jack Copeland, Toby Ord, Boris Pavlov and 
Andrew Rawlinson, who also helped me with the Figure 1.

\end{document}